\documentclass[%
reprint,
superscriptaddress,
preprintnumbers,
nobibnotes,
nofootinbib,
 amsmath,amssymb, 
 aps, prd,
 longbibliography,
]{revtex4-1}

\usepackage{cancel}
\usepackage{accents}
\usepackage{mciteplus,slashed}
\usepackage{amssymb,cancel,amsmath}
\usepackage{dcolumn}
\usepackage{bm}
\usepackage{soul}
\usepackage[caption=false]{subfig}
\usepackage{appendix}
\usepackage{physics}
\usepackage{booktabs}
\usepackage{comment}
\usepackage[T1]{fontenc}	
\usepackage{csvsimple}
\usepackage[bookmarksnumbered=true]{hyperref} 
\hypersetup{
    colorlinks = true,
    linkcolor = blue,
    anchorcolor = blue,
    citecolor = blue,
    filecolor = blue,
    urlcolor = blue
    }
\usepackage[section]{placeins}
\usepackage[capitalise]{cleveref}
\usepackage{booktabs}
\usepackage{graphicx}
\usepackage{mathrsfs}
\graphicspath{{Figures/}}
\AtBeginDocument{
\heavyrulewidth=.08em
\lightrulewidth=.05em
\cmidrulewidth=.03em
\belowrulesep=.65ex
\belowbottomsep=0pt
\aboverulesep=.4ex
\abovetopsep=0pt
\cmidrulesep=\doublerulesep
\cmidrulekern=.5em
\defaultaddspace=.5em
}
\usepackage[dvipsnames]{xcolor}
\usepackage[normalem]{ulem}
\usepackage{fontawesome} 
\usepackage[force]{feynmp-auto}
\usepackage{bbm}
\usepackage{MnSymbol}
\usepackage{comment}

\def\iu{\mathrm{i}}

\def\hyd{~\!\!^1{\rm H}}

\definecolor{forestgreen}{HTML}{228B22}

\begin{document}

\title{Atomic binding corrections for high energy fixed target experiments }

\author{Ryan Plestid}
\author{Mark B.\ Wise}
\affiliation{Walter Burke Institute for Theoretical Physics, California Institute of Technology, Pasadena, CA 91125, USA}

\date{\today}

\preprint{CALT-TH/2024-010}

\begin{abstract}
    High energy beams incident on a fixed target may scatter against atomic electrons. To a first approximation, one can treat these electrons as free and at rest. For precision experiments, however, it is important to be able to estimate the size of, and when necessary calculate, sub-leading corrections.  We discuss atomic binding corrections to relativistic lepton-electron scattering. We analyze hydrogen in detail, before generalizing our analysis to multi-electron atoms. Using the virial theorem, and many-body sum rules, we find that the corrections can be reduced to measured binding energies, and the expectation value of a single one-body operator. We comment on the phenomenological impact for neutrino flux normalization and an extraction of hadronic vacuum polarization from elastic muon electron scattering at MUonE.
\end{abstract}

 \maketitle

\section{Introduction}
Leptonic interactions provide a clean laboratory for precision physics.  In this work we consider high energy leptons (specifically neutrinos and muons) scattering on atomic electrons in a fixed target. 
When considering fixed target experiments the asymptotic in-state includes atomic electrons bound to nuclei. For certain  experiments, high levels of precision are necessary. Precision goals can range from modest to extremely demanding. As an example of a modest goal, the future DUNE experiment will aim for a percent-level determination of their neutrino flux normalization using $\nu e \rightarrow \nu e$ scattering \cite{DUNE:2021tad}. The Moller collaboration will attempt a sub-percent measurement of parity violation using an $11~{\rm GeV}$ beam of electrons scattering off atomic electrons in a hydrogen target \cite{MOLLER:2014iki}.  A more stringent example is provided by the MUonE experiment \cite{Abbiendi:2016xup}, which aims to extract hadronic vacuum polarization from $\mu e \rightarrow \mu e$ scattering \cite{CarloniCalame:2015obs} with an error budget for the differential cross section's shape $(1/\sigma)\times \dd \sigma/\dd t$ at the level of 10 ppm \cite{Abbiendi:2022oks,Banerjee:2020tdt}. 

It is intuitively obvious that these systems are well approximated by treating the atomic electrons as at rest, and working in the lab frame. For experiments with demanding precision goals, approximating the electrons as free and at rest may not be sufficient at which point binding corrections should be included. At bare minimum a parametric estimate for the size of these corrections should be established, and if necessary they should be computed. A number of basic questions immediately arise: what controls the binding expansion? how large are these corrections? how do they scale with atomic number? how does one include relativistic corrections?

To answer these questions concretely, we focus on neutrino and muon scattering on atomic electrons. For example, consider neutrinos scattering from a hydrogen atom at rest (i.e., with zero momentum),
\begin{equation}
    \nu(\vb{k}) + \hyd(\vb*{0})  \rightarrow \nu(\vb{k}') +  e^-(\vb{p}')  + p^+(\vb{h}')~. 
\end{equation}
For a generic atom $A$ we consider $\nu A \rightarrow \nu e^- B^+$ and sum over all final states $B^+$. We focus on kinematics such that  $|\vb{p}'| \sim |\vb{k}'| \sim |\vb{k}|$ are large, and $|\vb{h}'| \sim \sqrt{2m_e \epsilon_A}$ is small, where $\epsilon_A$ is the binding energy of the atom; for hydrogen $\epsilon_H\simeq \alpha^2 m_e/2$. Our goal is to provide a simple universal formula for the binding corrections to the cross section. 
 
A natural question is what controls the size of binding corrections parametrically. For example when matrix elements depend on the Mandelstam variable $s=(k+p)^2$ with $k_\mu=(\omega, 0,0,\omega)$, considering an electron at rest vs.~in motion we find $s=m_e^2+2m_e \omega$ vs.\ $s=m_e^2 + 2 (E_e \omega -p_e \omega\cos\theta)$. This would suggest corrections of the size $p_e/m_e \sim \sqrt{\epsilon_A/m_e}$. Alternatively, one may guess that since energy conservation is modified by the binding energy $\epsilon_A$, and the available energy is of order $\omega$ that corrections will be parametrically given by $\epsilon_A/\omega$ a {\it dramatically} smaller correction. As we will see in what follows, the leading order corrections are of order $\epsilon_A/m_e$. 

We study these binding corrections in detail for $\nu e \rightarrow \nu e$ scattering from an argon atom, and for $\mu e \rightarrow \mu e$ scattering from a carbon atom. We find that the parametric scaling mentioned above holds in general, but that a accidental cancellation makes the binding correction to $\nu e \rightarrow \nu e$ scattering very small. By way of contrast, for $\mu e \rightarrow \mu e$ scattering the corrections are of their natural size. For MUonE, we find that these corrections are roughly $5\times$ larger than their stated error budget goal, and must be included in future analyses. We do not include final state interactions (which can enter at the same order in the expansion) deferring their treatment to future work \cite{Plestid_Wise_2}.

Our analysis differs qualitatively from non-relativistic processes such as dark matter electron scattering \cite{Kopp:2009et,Roberts:2016xfw,Essig:2017kqs}.  We always consider an ultra-relativsitic final state electron, and a precision target that is much more stringent than what is required for dark matter direct detection. To retain theoretical control, it is therefore essential to expand around a free electron at rest as the zeroth order approximation. Our approach also sums over all final states (including ionized debris), and does not rely on a mean-field model of the atom.



The rest of the paper is organized along the following lines. First in \cref{sec:kinematics} we define, without approximation, the kinematics and phase space for scattering from hydrogen. Next in \cref{sec:plane-waves} we describe how, at lowest order in the fine structure constant $\alpha$ (i.e.\ neglecting final state interactions), the relevant matrix element in the case of hydrogen can be related to well known free electron matrix elements. Next, in \cref{sec:weak-binding} we outline a consistent expansion scheme to compute binding corrections. In \cref{sec:beyond-H} we discuss realistic multi-electron atoms and how to use sum rules and the virial theorem to reduce the problem to the atomic matrix element of a single one-body operator. Next in \cref{sec:applications} we discuss applications at high energy fixed target experiments including DUNE and MUonE. Finally in \cref{sec:discussion} we summarize our findings and outline potentially interesting future directions.

\section{Kinematics and phase space \label{sec:kinematics} }
Consider a particle scattering on hydrogen producing a proton and electron in the final state. For definiteness, we will take this to be a neutrino. The relevant matrix element is written in terms of the weak Hamiltonian density and is given by, 
\begin{equation}
    \label{mel-def}
    \begin{split}
    T &= \int \dd^4 x \mel{\nu e^- p^+}{\mathcal{H}_W(x)}{\hyd \nu}~\\
    &= (2\pi)^4\delta^{(4)}(\Sigma \tilde{p}) \mel{\nu e^- p^+}{\mathcal{H}_W}{\hyd \nu}~.
    \end{split}
\end{equation}
The tilded energy and momentum conservation reads 
\begin{align}
    \label{e-cons}
    \Sigma \tilde{E} &= m_e-\epsilon_H + \omega - \omega' -E' -T_p'~\\
    \Sigma \tilde{\vb{p}} &= \vb{k}+(-\vb{h}') -\vb{k}'-\vb{p}'~,
    \label{p-cons}
\end{align}
where $T_p'$ is the proton's kinetic energy, $\omega$ is the incident neutrino's energy, $\omega'$ the outgoing neutrino's energy, $E'$ the outgoing electron's energy, and $\epsilon_H$ the binding energy of hydrogen. Notice that \cref{p-cons} is the same as we would obtain for a free electron in the initial state with momentum $\vb{p}=(-\vb{h}')$, but that \cref{e-cons} differs by the binding energy of hydrogen.

We will focus on scattering from an unpolarized target. We average over the orientation of all atoms, which is equivalent to averaging over the quantum number $m_J$ (i.e., the magnetic quantum number for total angular momentum). This gives the same spin-averaged matrix elements as would be obtained by summing over free-particle spin and averaging over $m_\ell$ (i.e., the orbital quantum number) \cite{Bechler_1993}. 
Then, suppressing all spin indices for simplicity, the cross section is given by 
\begin{equation}
    \label{main-phase-space}
   \sigma = \frac{1}{2 \omega} \frac{1}{2M_{{\rm H}}}\int \dd \Pi_\nu  \dd \Pi_e \dd \Pi_p ~(2\pi)^4 \delta^{(4)}(\Sigma\tilde{p})~ |\mathcal{M}|^2 ~,
\end{equation}
where $T =(2\pi)^4 \delta^{(4)}(\Sigma \tilde{p}) \mathcal{M}$ and Lorentz invariant phase space is written for a particle of species-$a$ as $\dd \Pi_a= \dd^3 p_a/[(2E_a) (2\pi)^3]$; this formula assumes a relativistic normalization of states. It is convenient to eliminate the neutrino's phase space using the 3-momentum delta function, 
\begin{equation}
   \sigma = \frac{1}{2 \omega} \frac{1}{2M_{{\rm H}}} \int \dd \Pi_p  ~\int \dd \Pi_e \frac{1}{2\omega'} ~(2\pi) \delta(\Sigma \tilde{E}) ~ |\mathcal{M}|^2 ~.
\end{equation}
The energy conserving delta function can be used to eliminate the angle of the outgoing electron, $\cos\theta'$, relative to the beam axis. The explicit solution is given by 
\begin{widetext}
\begin{equation}
    \label{cos-theta-prime-sol}
    \cos\theta'= \frac{2 \omega  (E'-m_e+|\vb{h}'|\cos\theta+\epsilon_H-T_p)-(E'-m_e+\epsilon_H-T_p)^2+\vb{h}'^2-2|\vb{h}'||\vb{p}'|\sin\phi \sin\theta'\sin\theta +\vb{p}'^2}{2 |\vb{p}'|(\omega + |\vb{h}'|\cos\theta)}~,
\end{equation}
\end{widetext}
where $\theta$ and $\phi$ label the directions of the outgoing proton. 
Including the relevant Jacobian and integrating over the azimuthal angle (trivially by symmetry) we then arrive at 
\begin{equation}
    \begin{split}
    \sigma = \frac{1}{2 \omega} \frac{1}{2M_{{\rm H}}}\int \dd \Pi_p  & \qty(1+ \frac{|\vb{h}'|\cos\theta}{\omega} )^{-1} \\
    &\hspace{0.1\linewidth}\int \frac{\dd E'}{8\pi \omega}  \qty(1+\frac{T_p}{\omega'})~ |\mathcal{M}|^2 ~.
    \end{split}
\end{equation}
The limits of integration run from $E_e=m_e$ (the threshold for ionization) to $E_e^{\rm max}$ which is given by solving
\begin{equation}
    E' + \sqrt{\vb{q}^2 + 2\vb{p}'\cdot \vb{q} + \vb{p}'^2} +T_p'= m_e-\epsilon_H+\omega ~.
    \label{e-cons-expl}
\end{equation}
where $\vb{q}=\vb{k}-\vb{h}'$. At this stage all of our expressions are exact, and we now begin to discuss sensible approximations. 

An important input in our analysis is that $\mathcal{M}$ is the matrix element involving a bound atomic electron. As we will see below, this implies that the outgoing proton has momentum on the order of $\alpha m_e$ (for hydrogen). Large proton momenta are suppressed by the atomic wavefunction. Consequently, the proton's kinetic energy $T_p$ can be counted as $T_p\sim \alpha^2 m_e^2/m_p$ and can be neglected relative to $\omega\sim \omega' \sim E'$. 

Naively, the relative corrections from the modified energy conservation conditions are suppressed by $\epsilon_H/\omega$. A careful analysis shows that this naive expectation holds true for the phase space, but {\it does not} hold true for the matrix element squared, $|\mathcal{M}|^2$. As we discuss near \cref{spin-avg-2}, conservation of energy and momentum can cause certain Lorentz invariant products to be much smaller than the size of their individual components in the lab frame. For example $p'\cdot k' \sim p\cdot k \sim m_e \omega$. This is much smaller than the naive estimate of $p'\cdot k' \sim \omega^2$. For this reason, we keep only the leading terms in the phase space but expand the matrix element to $O(\vb{p}^2/m_e^2)$ as discussed in detail in \cref{sec:weak-binding}.  

We drop all terms suppressed by $\lambda\sim \epsilon_H/m_p, \epsilon_H/\omega$ (for example the proton's recoil energy).\footnote{Terms that go like $\vb{h}'/\omega \sim \lambda^{1/2}$  vanish upon integration over the final-state proton's phase space for a rotationally invariant state.} Then, 
\begin{equation}
    \begin{split}
    \sigma = \frac{1}{2 \omega} \frac{1}{2M_{{\rm H}}}\int \dd \Pi_p \int \frac{\dd E'}{8\pi \omega} ~ |\mathcal{M}|^2 + O\qty(\lambda)~,
    \end{split}
\end{equation}
with 
\begin{equation}
    \begin{split}
    \cos\theta'&= \frac{2 \omega  (E'-m_e)-(E'-m_e)^2+\vb{p}'^2}{2 |\vb{p}'|\omega } +  O\qty(\lambda)~,
    \end{split}
\end{equation}
and 
\begin{equation}
    E_{\rm max}= \frac{1}{2} \left(\frac{m_e^2}{2 \omega +m_e}+2 \omega +m_e \right)~+  O\qty(\lambda)~.
\end{equation}
The binding corrections to final-state kinematics (e.g., the limits of integration over final electron energies) are suppressed by the large energy scales in the problem. They produce an effect of $O(\epsilon_H/\omega)$, which for $\omega \sim ~{\rm GeV}$ and $\epsilon_H \sim 10~{\rm eV}$ amounts to a $10^{-8}$ correction; this is much smaller than the shifts in the matrix element that we discuss below.

\section{Reduction to plane waves \label{sec:plane-waves} }
The matrix element of the weak Hamiltonian in \cref{mel-def} can be written in terms of plane-wave 
(i.e., free-electron) states. For the outgoing scattering state, this is trivial at lowest order in $\alpha$ since $\ket{e^-p^+}_{\rm out} \simeq \ket{e^-p^+}_{\rm PW}$ with the subscript denoting a plane-wave state. Final state interactions can be added perturbatively order-by-order in $\alpha$ but are not included in what we present below, but are studied separately in \cite{Plestid_Wise_2}. The bound states demand the inclusion of the Coulomb interaction at zeroth order, but may still be expanded in plane wave states. At leading order in $\alpha$ a hydrogen atom at rest can be written as\footnote{The state $\ket{e^-(\vb{p})}$ has a polarization spinor $u(\vb{p})$. This accounts for the leading  relativistic correction to $\psi(\vb{p})$ (see \cref{app:rel-corr} for a discussion). } \textcolor{red}{\cite{Bethe:1957ncq}}
\begin{equation}
    \label{hyd-exp}
    \ket{\hyd} \simeq \sqrt{2M_H}\int \frac{\dd^3 p}{(2\pi)^3} 
    \frac{\tilde{\psi}(\vb{p})}{\sqrt{2E_e}\sqrt{2 E_p}}~
    \ket{p^+e^-}_{{\rm PW}}~,
\end{equation}
with the proton carrying momentum $-\vb{p}$ and the electron carrying momentum $\vb{p}$. The approximation holds up to $O(\epsilon_H/m_e)$.

We suppress spin indices for simplicity anticipating spin averaging in the initial state. The function $\tilde{\psi}(\vb{p})$ is the Fourier transform of the coordinate space wavefunction $\psi(\vb{x})$. The factors in square roots are necessary when working with relativistically normalized states. The energies that appear are those of free particles,  $E_e=\sqrt{\vb{p}^2+m_e^2}$ and $E_p=\sqrt{\vb{p}^2+m_p^2}$. One can easily check that provided $\int \dd^3p/(2\pi)^3  |\tilde{\psi}(\vb{p})|^2 = 1$ and the electron and proton states are normalized relativistically, then it follows the hydrogen atom has the correct normalization also.

Using $\braket{p^+(\vb{h}')}{p^+(-\vb{p})} = 2 E_p(\vb{p}) (2\pi)^3 \delta^{(3)}(\vb{p}+\vb{h}')$, we may therefore write, to leading order in $\alpha$, 
\begin{equation}
    \begin{split}
    \mathcal{M} &=  \mel{\nu e^- p^+}{\mathcal{H}_W}{\hyd \nu} \\
                &\simeq \sqrt{ \frac{2 M_H 2 E_p}{2 E_e}} \tilde{\psi}(\vb{p}) 
       ~ {\sf M} (\vb{p}',\vb{k}',\vb{p},\vb{k})
    \end{split}
\end{equation}

~\\

\noindent where we have introduced the matrix element for $2\rightarrow 2$ scattering of an incident neutrino from a free electron with momentum $\vb{p}$,
\begin{equation}
    {\sf M}(\vb{p}',\vb{k}',\vb{p},\vb{k}) = \bra{\nu(\vb{k}') e^-(\vb{p}')}{\mathcal{H}_W}\ket{\nu(\vb{k}) e^-(\vb{p})}_{\rm PW}~.
\end{equation}
Note that this matrix element is a function of the three momenta only. 

Writing the cross section in terms of the free-electron matrix element we then find, 
\begin{equation}
    \begin{split}
    \sigma \textcolor{red}{\simeq}  \int \frac{\dd^3 p}{(2\pi)^3} |\tilde{\psi}(\vb{p})|^2~ \frac{1}{2E}\frac{1}{2 \omega} \int \frac{\dd E'}{8\pi \omega} ~ \frac12\sum_{\rm spins} |{\sf M}|^2 ~,
    \end{split}
\end{equation}
where spin sums and averages which were previously left implicit have been made explicit. This formula holds up to an accuracy of $O(\epsilon_A/m_e)$. The factor of $1/2$ in this formula results from averaging over the initial electron spin and assumes the incident neutrino is polarized (as it must be). 

\begin{widetext}
Let us now evaluate the free-electron matrix element for the example at hand of $\nu e \rightarrow \nu e $ scattering using 
\begin{equation}
    \begin{split}
    {\sf M} = -\sqrt{8}G_F &\qty[\bar{u}(\vb{k}') \gamma^\mu P_L u(\vb{k})] \qty[ \bar{u}(\vb{p}') \gamma^\mu (c_L P_L + c_R P_R) u(\vb{p})]~,
    \end{split}
\end{equation}
$P_{L,R}=(1\mp \gamma_5)/2$ are chiral-projectors. Summing over spins and performing traces over Dirac matrices we obtain 
\begin{equation}
    \label{spin-avg-2}
      \sum_{\rm spins} |{\sf M}|^2= 128 G_F^2
      \times \big[c_L^2 (p\cdot k)(p'\cdot k')+ c_R^2 (p\cdot k')(k\cdot p') -4 c_L c_R m_e^2 (k\cdot k')\big] ~.
\end{equation}
\end{widetext}
We have used the notation $k\cdot p = k^\mu p_\mu=\omega\sqrt{\vb{p}^2+m_e^2}  - \vb{k}\cdot \vb{p}$; these invariants are only functions of three-momenta.

Notice that the matrix element squared is of order $|{\sf M}|^2 \sim m_e^2 \omega^2$. To see this one can use energy and momentum conservation to relate Lorentz invariants involving two four-vectors with large energy components such as $p'\cdot k'$ to Lorentz invariants involving $p$ such as $p\cdot k$. After this has been done it becomes immediately apparent that the matrix element is $O(m_e^2 \omega^2)$ with a series of corrections controlled by the ratio $|\vb{p}|/m_e$.

At this point it is important to emphasize that the kinematic relations for our process  differ from those of $2\rightarrow 2$ electron scattering. In particular when using energy and momentum conservation to relate $(p'\cdot k')$ to $(p\cdot k)$ we must account for the binding energy. Let us define $\langle V \rangle \equiv m_e-\epsilon_H - E(\vb{p})$, and introduce the four-velocity of the hydrogen atom $u_\mu=(1,0,0,0)$ in the lab frame. We then have
\begin{equation}
    p_\mu + k_\mu + \langle V \rangle u_\mu = p'_\mu + k'_\mu ~. 
\end{equation}
This allows us to trade different Lorentz invariants for one another, 
\begin{align}
    \begin{split}
    \label{kin-rel-1}
    p'\cdot k' &= p\cdot k + \langle V \rangle (p+k)\cdot u  + \tfrac12\langle V \rangle^2\\
    &\simeq p\cdot k + \langle V \rangle \omega ~,
    \end{split}\\
    \begin{split}
    \label{kin-rel-2}
    p' \cdot k &= p\cdot k' + \langle V \rangle (k'-p)\cdot u  -  \tfrac12 \langle V \rangle^2~\\
    &\simeq p\cdot k' + \langle V \rangle \omega'  ~.
    \end{split}\\
    \begin{split}
    \label{kin-rel-3}
    k \cdot k' &= p\cdot p'+m_\ell^2-m_e^2 + \langle V \rangle (p'-p)\cdot u  - \tfrac12\langle V \rangle^2~\\
    &\simeq p\cdot p'+m_\ell^2 + \langle V \rangle E'~,
    \end{split}
\end{align}
where $m_\ell=0$ for $\nu e \rightarrow \nu e$ scattering, but will be equal to the muon mass when we consider $\mu e \rightarrow \mu e$ scattering. 

\section{Weak binding expansion \label{sec:weak-binding} }
In the limit of weak binding we may estimate $|\vb{p}| \sim  \sqrt{2 \epsilon_H m_e}$ and expand in $\epsilon_H/m_e$. In what follows, the use of the symbol ``$\simeq$'' denotes results that hold up to and including $O(\epsilon_H/m_e)$ in the weak binding expansion.
We need only the first terms in $\vb{p}^2$ and $\langle V \rangle$ separately. Setting $m_\ell^2=0$, 
\begin{align}
    (p\cdot k)(p'\cdot k') &\simeq (p\cdot k)(p\cdot k+\omega \langle V \rangle) \\
    &\simeq (p\cdot k)_0^2 \qty[1 + \frac{\vb{p}^2}{m_e^2}(1+\cos^2\theta_{e\nu}) + \frac{\langle V \rangle}{m_e}]~, \nonumber \\[6pt]
    (p\cdot k')(p'\cdot k) &\simeq (p\cdot k')(p\cdot k'+\omega' \langle V \rangle) \\
    &\simeq (p\cdot k')_0^2 \qty[1 + \frac{\vb{p}^2}{m_e^2}(1+\cos^2\theta_{e\nu'}) + \frac{\langle V \rangle}{m_e}]~, \nonumber \\[6pt]
    k\cdot k' &\simeq p\cdot p'+\omega' \langle V \rangle  \\
    &\simeq (p\cdot p')_0 \qty[1 + \frac{\vb{p}^2}{2m_e^2} + \frac{\langle V \rangle}{m_e}]~,\nonumber 
\end{align}
where we have dropped terms linear in $\vb{p}$ that vanish upon averaging over the electron's wavefunction. 
The outgoing electron and neutrino are nearly aligned with the beam axis, with angles set by $1/\gamma \sim \sqrt{m_e/\omega}$ where $\gamma$ is the boost to the center of mass frame. 
We may therefore set $\cos\theta_{e\nu'} \simeq \cos\theta_{ee'} \simeq \cos\theta_{e\nu} \equiv \cos\theta$.\!\footnote{If one worked with e.g., $k\cdot k'$ rather than $p\cdot p'$, then large cancellations occur $\omega \omega' - \vb{k} \cdot \vb{k}'\sim O(m_e \omega)$. One cannot set $\cos\theta_{e\nu'} \simeq \cos\theta_{ee'} \simeq \cos\theta_{e\nu}$ and must retain higher order terms.}

The subscript zero denotes the Lorentz invariant evaluated for an elecron at rest e.g., $(p\cdot k)^2_0=(m_e^2 + 2 m_e \omega)^2$. We observe that the terms multiplying $c_L^2$ and $c_R^2$ in \cref{spin-avg-2} will be multiplied by a common factor, whereas the term proportional to $c_L c_R$ is not. Nevertheless this cross term is lepton-mass suppressed ($m_e/\omega \sim 10^{-3}$ for $\omega \sim 1~{\rm GeV}$), and we can neglect it when computing the corrections.

The cross section can now be written in the simple form
\begin{equation}\label{Hyd-sigma}
    \begin{split}
    \sigma \simeq & \int \frac{\dd^3 p}{(2\pi)^3} |\tilde{\psi}(\vb{p})|^2\qty(1+\frac{1}{3}\frac{\vb{p}^2}{m_e^2} -\frac{\epsilon_H}{m_e} ) \\
    &\hspace{0.25\linewidth}\times \frac{1}{2m}\frac{1}{2 \omega} \int \frac{\dd E'}{8\pi \omega} ~ \frac12\sum_{\rm spins} |{\sf M}|_0^2 ~.
    \end{split}
\end{equation}
where we have expanded $1/E = 1/m(1-\vb{p}^2/2m_e)$.
Notice that the corrections begin at $O(\vb{p}^2/m_e^2)$ much like in the case of bound-muon decays \cite{
PhysRevD.61.073001,PhysRev.119.365,PhysRev.120.1450}. We recognize the second line as the cross section for scattering with an electron at rest. We have exchanged $\langle V \rangle$ for $\epsilon_H$ using 
$\vb{p}^2/2m_e+ \langle V \rangle = -\epsilon_H$. We define the binding correction  $\dd \sigma_{\mathcal{B}}$ via $\dd \sigma \simeq \dd \sigma_0 + \dd \sigma_{\mathcal{B}}$. We therefore find at the level of the differential cross section, the binding correction can be written as 
\begin{equation}
    \begin{split}
    \dd \sigma_{\mathcal{B}} &= \frac{1}{m_e}\frac{\mel{\hyd}{ \tfrac23 \hat{T} + \hat{H}}{\hyd}}{\braket{\hyd}{\hyd}}\times \dd \sigma_0 \\
    &=  \frac{1}{m_e}\langle \tfrac23\hat{T} + \hat{H} \rangle \dd \sigma_0 ~, 
    \end{split}
\end{equation}
\bigskip

\noindent where $\langle \ldots \rangle$ means expectation value divided by the states norm.
This simple factorized form is a consequence of the symmetric nature of the leading terms proportional to $c_L^2$ and $c_R^2$ in the tree-level matrix element. Note that $\langle \hat{T} \rangle $  is related to $\langle \hat{H} \rangle$ by the virial theorem, and so the entire correction is fixed in terms of hydrogen's binding energy.  

\section{Beyond hydrogen \label{sec:beyond-H} }
When considering more complicated atoms the analysis above must be generalized. We are in fact interested in the sum over all final atomic states. Let us consider a reaction 
\begin{equation}
    \nu(\vb{k}) + A(\vb*{0}) \rightarrow \nu(\vb{k}') + e^-(\vb{p}') + B^+(\vb{h}')~,
\end{equation}
where $B$ is an atomic system with one fewer electron than $A$, with center of mass momentum $\vb{h}'$,  but otherwise unconstrained. We can separate out the final state phase space of $B^+$ into its center of mass motion and ``the rest'' $\sum_B$. We therefore have 
\begin{equation}
    \begin{split}
    \sigma = \frac{1}{2 M_A} \frac{1}{2\omega} \sum_B\int \dd \Pi_B  &\int \dd \Pi_\nu  \dd \Pi_e ~\\
    &~~(2\pi)^4 \delta^{(4)}(\Sigma\tilde{p})~ |\mathcal{M}_B|^2~.
    \end{split}
\end{equation}
\vspace{12pt}

\noindent The energy conserving delta function now carries a dependence on the final state $B$ 
\begin{equation}
    \begin{split}
    \Sigma \tilde{E} &= E_A+ \omega - E' - \omega' - E_B~.
    \end{split}
\end{equation}
As above, we neglect the recoil energy of the system $B$ 
\begin{equation}
    E_A-E_B \simeq (m_e - \epsilon_A) + \epsilon_B~,
\end{equation}
where $\epsilon_A$ and $\epsilon_B$ are the atomic binding energies of $A$ and $B^+$ respectively. Introducing %
\begin{equation}
    {\cal E}=m_e+\omega -(E'+\omega')~,
\end{equation}
(the energy conservation condition for an electron at rest) we have
\begin{equation}
    \Sigma \tilde{E} = {\cal E}- (\epsilon_A-\epsilon_B)~. 
\end{equation}

Using the mode expansion of the field $\psi_e$, and again neglecting final state interactions, the matrix element can be written in the form
\begin{equation}
    \begin{split}
    \mel{e^- B^+}{\bar{\psi}_e \Gamma \psi_e}{A} = \int \frac{\dd^3 q}{(2\pi)^3\sqrt{2E_q}}\bar{u}&(\vb{p}') \Gamma u(\vb{q}) \\
    &\times \mel{B^+}{\hat{a}_{\vb{q}} }{A}~. 
    \end{split}
\end{equation}
 with $\Gamma$ an appropriate Dirac bilinear. Notice that $\mel{B^+(\vb{h}')}{\hat{a}_{\vb{q}} }{A(\vb*{0})} \propto (2\pi)^3 \delta^{(3)}(\vb{q}-\vb{h}')$. We can therefore replace $u(\vb{q)}\leftrightarrow u(\vb{h}')$. It is convenient to insert $\int \dd \epsilon ~\delta(\epsilon-\epsilon_A+\epsilon_B)$ and such that $\sum_B$ can be pulled through the other integrals, 

\begin{widetext}
\begin{equation}
    \label{long-eq-sigma}
    \begin{split}
   \sigma =  \int \dd \epsilon   &\int \frac{1}{2 \omega} \int  \frac{\dd \Pi_e}{2\omega'}(2\pi) \delta({\cal E}-\epsilon) 
     \\
   & \sum_B  \int \dd \Pi_B\int \frac{\dd^3 q}{(2\pi)^3 \sqrt{2E_{q}}}\frac{\dd^3 p}{(2\pi)^3 \sqrt{2E_{p}}}\delta(\epsilon_A-\epsilon_B-\epsilon) ~ \mel{A}{\hat{a}_{\vb{q}}^\dagger}{B^+}\mel{B^+}{\hat{a}_{\vb{p}} }{A} ~\frac12  \sum_{\rm spins}   |{\sf M}(\vb{p},\vb{k},\vb{p}',\vb{k}')|^2~,
   \end{split}
\end{equation}
\end{widetext}
where the free-electron matrix element, ${\sf M}$ is defined above. We have anticipated that the integrand in \cref{long-eq-sigma} is proportional to $\delta^{(3)}(\vb{p}-\vb{q})$, and that the atomic matrix elements are  spin diagonal after averaging over initial spins of $A$ and summing over the finals spins of $B$. 
\Cref{long-eq-sigma} may be conveniently re-written in terms of the spectral function using $\sumint_B= \sum_B \int \dd \Pi_B$, 
\begin{align}
    S_A(\epsilon,\vb{p}) =& \int \frac{\dd^3 q}{(2\pi)^3} \sumint_B \mel{A}{\hat{a}^\dagger_{\vb{p}}}{B} \delta(\epsilon_A-\epsilon_B-\epsilon) \mel{B}{\hat{a}_{\vb{q}}}{A}\nonumber \\
    =&\int \frac{\dd^3 q}{(2\pi)^3}\sumint_B \mel{A}{\hat{a}^\dagger_{\vb{p}} \delta(\epsilon_A\textcolor{red}{+}\hat{H}-\epsilon)}{B} \mel{B}{\hat{a}_{\vb{q}}}{A} \nonumber \\
    =&\int \frac{\dd^3 q}{(2\pi)^3}\mel{A}{\hat{a}^\dagger_{\vb{p}}\delta(\epsilon_A\textcolor{red}{+}\hat{H}-\epsilon) \hat{a}_{\vb{q}} }{A}~. 
\end{align}
The Hamiltonian $\hat{H}$ is defined with all rest masses of elementary particles subtracted. It may be written as $\hat{H}= \hat{T}+\hat{V}_1 + \hat{V}_2$ i.e., as a kinetic energy, electron-nucleus potential, and inter-electron potential contribution. We then have 
\begin{align}
   \sigma &=  \int \dd \epsilon \int \frac{\dd^3 p}{(2\pi)^3 } S_A(\epsilon,\vb{p})\\
   &\hspace{0.15\linewidth}\frac{1}{2 \omega} \frac{1}{2E}\int \dd \Pi_e \frac{1}{2\omega'} \delta({\cal E}-\epsilon)\frac12\sum_{\rm spins} |{\sf M}|^2 .\nonumber
\end{align}
The energy and momentum conservation conditions now read
\begin{equation}
    E+\omega + \tilde{V} = E'+\omega'~,
\end{equation}
with $\tilde{V}\simeq  -\epsilon - \vb{p}^2/2m_e$. We then find for the cross section
\begin{equation}
    \begin{split}
    \sigma \simeq & \int \dd \epsilon\int\frac{\dd^3 p}{(2\pi)^3} S_A(\epsilon,\vb{p}) ~\qty(1+\frac{1}{3}\frac{\vb{p}^2}{m_e^2} -\frac{\epsilon}{m_e}) \\
    &\hspace{0.25\linewidth}\times \frac{1}{2m}\frac{1}{2 \omega} \int \frac{\dd E'}{8\pi \omega} ~ \frac12\sum_{\rm spins} |{\sf M}|_0^2 ~,
    \end{split}
\end{equation}
which holds to an accuracy of $\epsilon_A/m_e$. When computing the corrections from $\vb{p}^2$ and $\epsilon$, it is sufficient to use a non-relativistic spectral function which satisfies energy weighted sum rules \cite{Polls:1994zz} (the latter of which is due to Koltun \cite{Koltun:1972kh})
\begin{align}
    &\int \dd \epsilon\int \!\frac{\dd^3 p}{(2\pi)^3} \frac{\vb{p}^2}{2m_e}S_A(\epsilon,\vb{p})  = \langle \hat{T} \rangle_A = \epsilon_A ~,\\ 
    &\int \dd \epsilon\!\int \!\frac{\dd^3 p}{(2\pi)^3} (-\epsilon)~ S_A(\epsilon,\vb{p})  =  \langle \hat{T} \rangle_A + \langle \hat{V}_1 \rangle_A + 2 \langle \hat{V}_2 \rangle_A \nonumber \\
    &\hspace{0.5\linewidth}=\qty(-3\epsilon_A- \langle \hat{V}_1 \rangle_A)~.
\end{align}
To relate the matrix elements to binding energies we have used the virial theorem: $\langle T\rangle_A =  \epsilon_A$ and $\langle \hat{V}_1 + {V}_2 \rangle_A=-2\epsilon_A$. For a sketch derivation of the sum rules see \cref{app:sum-rules}.
We then find for the binding correction at order $\epsilon_A/m_e$ 
\begin{equation}
    \sigma_{\mathcal{B}} =  \frac{1}{m_e}\frac{\mel{A}{ \tfrac53\hat{ T}  +  \hat{V}_1 + 2 \hat{V}_{2}}{A} }{Z_A\braket{A}{A}}\sigma_0~.
\end{equation}
Therefore, at the level of differential cross section for $\nu e \rightarrow \nu e$ scattering we have 
\begin{equation}
    \label{main-nue}
    \begin{split}
    \dd \sigma_{\mathcal{B}} &=  \frac{1}{Z_A m_e} \qty(-\frac73 \epsilon_A  -  \langle \hat{V}_1 \rangle_A ) \dd \sigma_0~.
    \end{split}
\end{equation}
The relevant atomic input for the leading order binding corrections is therefore the measurable binding energy, and a single one-body matrix element, $\mel{A}{\hat{V}_1}{A}/\braket{A}{A}$. To evaluate this matrix element for an atom with $Z_A$ electrons the radial density field of electrons $n_A(\vb{r}) = \langle \hat{n}(\vb{r}) \rangle_A$ is sufficient
\begin{equation}
   \langle \hat{V}_1 \rangle_A = Z_A\times \int \dd^3 r ~~ n_A(\vb{r}) \qty(\frac{-Z_A\alpha}{|\vb{r}|})~,
\end{equation}
where $\int \dd^3 r ~n_A(\vb{r}) =1$. This is particularly interesting because $n_A(\vb{r})$ is amenable to calculation, and is in general easier to predict than e.g., the full many-body wavefunction.  

To estimate parametric scaling with $Z_A$ we can make use of the Thomas-Fermi model (see \cref{TF-model}), which should be accurate at the 10\% level or better.\!\footnote{More accurate many-body calculations and the Thomas-Fermi model differ by $\sim 10\%$ at the level of $n_A(\vb{r})$, but these differences tend to largely cancel for integrated quantities such as $\langle \hat{V}_1 \rangle_A$.} One finds,
\begin{equation}
    \label{thomas-fermi-est}
    \langle \hat{V}_1\rangle_A \approx -Z_A^{7/3}\times (30~{\rm eV})~.
\end{equation}
In this expression one factor of $Z_A$ comes from the total number of electrons in the atom. Another factor of $Z_A$ comes from the nuclear Coulomb potential. The final factor of $Z_A^{1/3}$ comes from the Thomas-Fermi length scale $b_0 \approx 0.88 Z_A^{-1/3} a_0$ with $a_0=1/(\alpha m_e)$ the Bohr radius. 

\section{Applications \label{sec:applications} }
In this section we discuss some simple applications of our methods to problems of physical interest. We focus DUNE and MUonE specifically. 
\subsection{Neutrino flux measurements}
An immediate application of our results is the measurement of neutrino flux normalization at future high intensity neutrino experiments. For concreteness we will focus on scattering event rates at the DUNE near detector with a liquid argon time projection chamber (LArTPC). This is interesting both because of DUNE's anticipated high statistical sample of $\nu e \rightarrow \nu e$ scattering, but also because argon is a noble element. Therefore, the treatment used above in terms of a spherically symmetric isolated atom (i.e., without complications due to molecular physics) applies immediately. 

Recent estimates suggest that a measurement of the $\nu e$ elastic scattering rate in a 30 ton LArTPC in the near detector hall can supply a measurement of the neutrino flux normalization with a  $2\%$ uncertainty after five years of operation \cite{Marshall:2019vdy}. The authors of Ref.~\cite{Marshall:2019vdy} assume a $\nu e$ elastic scattering rate of $\sim 120 \times 30 = 3.6\times 10^3$  events per year for a 30 ton liquid argon detector and a five year run time. The DUNE conceptual design report suggests roughly twice as large a statistical sample will be available per year \cite{DUNE:2021tad}. When combined with a ten vs. five year run time this could reduce statistical errors by a factor of two. Higher statistics, and improved control over detector systematics may then allow for determination of the neutrino flux with $\lesssim 1\%$ precision. It is important to have theoretical control over all possible corrections at a sub-percent level. Estimating the effect of binding corrections {\it a priori} one would expect a correction of size  $\sim Z_{\rm Ar}^{4/3} \alpha^2 \approx 2.5\times 10^{-3}$. 

\vspace{3pt}

 The total binding energy of argon  is given by $\epsilon_{\rm Ar} = 14.40~{\rm keV}$ \cite{NIST_ASD}.   
Estimating the average inverse radius using tabulated Hartree-Fock calculations from Ref.~\cite{osti_4553157}, and averaging over atomic orbitals, we find $\langle V_1\rangle_{\rm Ar} = -34.2~{\rm keV}$. 

We then find for the binding correction to the cross section for scattering on atomic electrons in argon, 
\begin{equation}
    \label{nu-pheno-est}
    \begin{split}
    \dd \sigma_{\cal B} &= \qty(- \qty[\frac{7}{3}\qty(1.57) -3.71  ] \times 10^{-3})\dd \sigma_0\\  
    &\approx (-6 \times10^{-5} ) \times \dd \sigma_0~. 
    \end{split}
\end{equation}
This is much smaller than the previously estimated error on the cross section ($\pm 0.37\%$) \cite{Tomalak:2019ibg}. We find due to the delicate cancellation between binding energy and one-body potential terms, that this estimate is very sensitive to the precise value of each term. Nevertheless, the size is always much smaller than $10^{-3}$ and can therefore be safely neglected for neutrino flux normalization measurements with an argon target.
\subsection{Muon electron scattering} 

MUonE will measure muon electron scattering with tagged incoming muons, and excellent angular resolution for outgoing electrons and muons. Bound state effects can be, and indeed are, important for MUonE. The purpose of the experiment is to measure the running of $\alpha(\mu)$ where $\alpha$ is the electromagnetic fine structure constant. The strategy may be summarized schematically as \cite{Banerjee:2020tdt}
\begin{equation}
    \label{rad-corr-muone}
    \frac{1}{\sigma}\dv{\sigma}{t} = \frac{1}{\sigma^{(0)}}\dv{\sigma^{0}}{t} \qty|\frac{\alpha(t)}{\alpha(0)}|^2 \times (1+ \delta_R(t))~, 
\end{equation}
where $\delta_R$ contains radiative corrections. Note that this method requires only the {\it shape} of the differential cross section and not its overall normalization. Since hadronic vacuum polarization is a small effect \cite{CarloniCalame:2015obs,Abbiendi:2016xup}, the demands for $\delta_R(t)$ are extraordinarily stringent (10 ppm) \cite{Banerjee:2020tdt}, as are the demands for systematic experimental uncertainties \cite{Abbiendi:2022oks}. The ultimate performance of the detector is beyond our control, but our goal is to help reduce theoretical errors stemming from bound-state corrections below the requisite target of 10 ppm.

As we have discussed above in the case of neutrino scattering, bound-state corrections should also be included i.e., one should make the replacement $(1+ \delta_R)\rightarrow (1+ \delta_R+\delta_{\cal B})$. As we will see, the bound state corrections to $\mu e \rightarrow \mu e $ have important phenomenological implications for an extraction of hadronic vacuum polarization at MUonE. 

The angular resolution at  MUonE is superb; $0.02~{\rm mrad}$ \cite{Abbiendi:2022oks}. It is therefore interesting to ask if the effects from atomic binding can significantly shift the kinematics of $\mu e \rightarrow \mu e$ scattering events. We have checked the modified kinematic constraints, and find that for the $\sim 150~{\rm GeV}$ muon beam at MUonE, the shifts in $\theta_\mu$ and $\theta_e$ are suppressed by $\epsilon_A/E_\mu \ll 0.02~{\rm mrad}$ and are therefore irrelevant for practical considerations.

At tree-level $\mu e\rightarrow \mu e$ is mediated by an off-shell photon, whose virtuality is given by $q^2=(k-k')^2$ with $k$ the four-momentum of the incoming muon, and $k'$ the momentum of the outgoing muon. The relevant matrix element in our problem is then $\mel{e^- B^+}{\hat{J}_\mu}{A}$ with $\hat{J}_\mu$ the (leptonic) electromagnetic current. Following identically to above, for the reaction
\begin{equation}
    \mu^\pm (\vb{k}) + A \rightarrow \mu^\pm(\vb{k}') + e^-(\vb{p}') + B(-\vb{p})~, 
\end{equation}
we arrive at 
\begin{equation}
    \begin{split}
    \sigma = & \int \dd \epsilon\int\frac{\dd^3 p}{(2\pi)^3} S_A(\epsilon,\vb{p})  \\
    &\hspace{0.25\linewidth}\times \frac{1}{2E}\frac{1}{2 |\vb{k}|} \int \frac{\dd E'}{8\pi \omega} ~ \frac14\sum_{\rm spins} |{\sf M}|^2 ~,
    \end{split}
\end{equation}
where the factor of $1/4$ comes from averaging over both electron and muons spins in the initial state. 
Neglecting $m_e/\omega$ and $m_e^2/m_\mu^2$, the spin-summed matrix element is given by 
\begin{equation*}
    \sum_{\rm spins} |{\sf M}|^2 \simeq 32e^4 \frac{(k\cdot p')(k'\cdot p) + (k\cdot p)(k'\cdot p')-m_\mu^2 (p\cdot p')}{\qty[(k-k')^2]^2} ~.
\end{equation*}
When expanding Lorentz invariants in terms of $\vb{p}$ we must be more careful than in the case of neutrino scattering. Terms proportional to $\cos\theta$ can appear from both the numerator and denominator. One should therefore keep terms linear in $\vb{p}$, and the appropriate identities are  
\begin{align}
    (p\cdot k)&\simeq (p\cdot k\phantom{'})_0 \qty[1 - \frac{|\vb{p}|\cos\theta}{m_e}+ \frac{\vb{p}^2}{2m_e^2}]~, \\
    (p\cdot k') &\simeq (p\cdot k')_0\qty[1 - \frac{|\vb{p}|\cos\theta}{m_e}+ \frac{\vb{p}^2}{2m_e^2}] ~,\\
    (p\cdot p') &\simeq (p\cdot p')_0\qty[1 - \frac{|\vb{p}|\cos\theta}{m_e}+ \frac{\vb{p}^2}{2m_e^2}]~,
\end{align}
where we (again) use $\cos\theta_{ee'}\simeq \cos\theta_{e\mu'} \simeq \cos\theta_{e\mu}\equiv \cos\theta$. Next, we expand in $\vb{p}^2$ and  $\epsilon$. After averaging over angles in the bound state, we may make the replacement, 

\begin{figure}
    \centering
    \includegraphics[width=\linewidth]{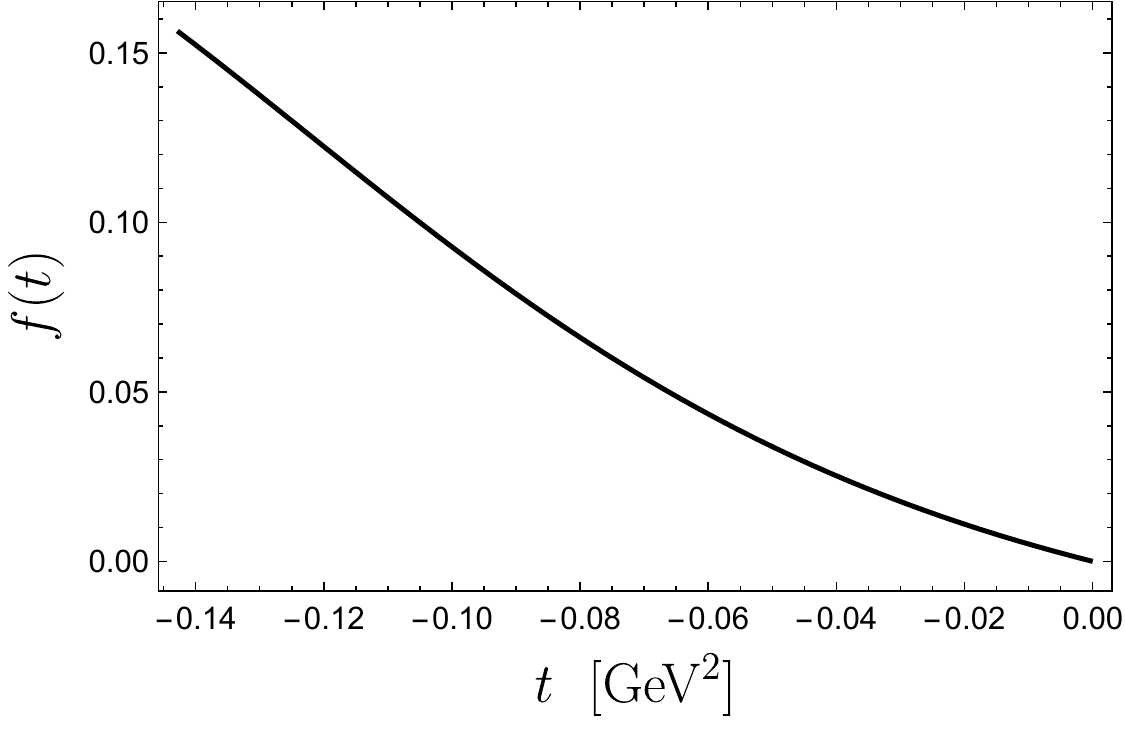}
    \caption{The function $f(t)$, as defined in \cref{f(t)}, for $\sqrt{s}=405~{\rm MeV}$, plotted  over the interval $t_{\rm min} \leq t \leq 0$ where the largest  momentum transfer is given by  $t_{\rm min} = - (s-m_\mu^2)^2/s$. 
    \label{fig:f(t)} }
\end{figure}
\begin{equation}
    \label{replacement}
    \begin{split}
   \frac{1}{2E} \sum_{\rm spins} |{\sf M}|^2 \rightarrow &\frac{1}{2m} \bigg\{ ~\sum_{\rm spins} |{\sf M}|^2_0 \qty[1+ \frac{\epsilon }{m_e} ] \\
   &\hspace{0.15\linewidth}- \frac{8 e^4 m_\mu^2}{(p\cdot p')_0} \qty[\frac{\epsilon}{m_e}+\frac13\frac{\vb{p}^2}{m_e^2} ] ~\bigg\}.
   \end{split}
\end{equation}
Upon integration against the spectral function and using the relevant sum rules, we find at $O(\epsilon_A/m_e)$, 
\begin{equation}
    \label{main-mue-a}
\begin{split}
    \dv{\sigma}{t} 
    &\simeq \dv{\sigma^{(0)}}{t} \bigg[ 1 + \frac{1}{Z_A m_e}\qty[3\epsilon_A   +  \langle \hat{V}_1 \rangle_A ]  \\
    &\hspace{0.3\linewidth}-   \frac{f(t)}{Z_Am_e}  \qty(\frac{11}{3}\epsilon_A  + \langle \hat{V}_1 \rangle_A)   \bigg]~,
\end{split}
\end{equation}
where the Mandelstam variable $t$ for an electron at rest is defined as $t\simeq-2(p\cdot p')_0=-2m_e E'$ in the limit of small electron mass. The function $f(t)$ (plotted in \cref{fig:f(t)} for $\sqrt{s}=405~{\rm MeV}$) is obtained from \cref{replacement} after dividing through by the leading order answer, 
\begin{align}
\label{f(t)}
    \begin{split}
    f(t)&= \frac{m_\mu^2 (p\cdot p')_0}{(p\cdot k)_0^2 + (p\cdot k')_0^2-m_\mu^2(p\cdot p')_0 } \\
    &=\frac{-2m_\mu^2 t}{(s+t-m_\mu^2)^2 + (s-m_\mu)^2+2m_\mu^2t } ~.
    \end{split}
\end{align}
For a shape-only measurement, the relevant quantity is 
\begin{equation}
\label{main-mue-b}
\begin{split}
    \frac{1}{\sigma}\dv{\sigma}{t} 
    &\simeq \frac{1}{\sigma^{(0)}}\dv{\sigma^{(0)}}{t} \bigg( 1 -\frac{f(t)}{Z_A m_e}\qty[\frac{11}{3}\epsilon_A   +  \langle \hat{V}_1 \rangle_A ]\bigg)~,
\end{split}
\end{equation}
To get a sense of the size of this correction we may assume a carbon target, taking $\epsilon_{\rm C} =1.03~{\rm keV}$ \cite{NIST_ASD} and $\langle V_1 \rangle_{\rm C}=-2.40~{\rm keV}$ \cite{osti_4553157}. We find, 
\begin{equation}
    \label{mu-pheno-est}
    \frac{1}{Z_A m_e}\qty[\frac{11}{3}\epsilon_A   +  \langle \hat{V}_1 \rangle_A] \simeq 45 \times 10^{-5}~. 
\end{equation}
This is almost two orders of magnitude larger than the theory-error target of MUonE. Using $\sqrt{s}=405~{\rm MeV}$ and $\sqrt{-t}=\sqrt{-t_{\rm peak}} = 330~{\rm MeV}$ \cite{Abbiendi:2022oks} we find that this correction amounts to a $5\times 10^{-5}$ shift which is a sizeable effect,  when compared to the $10^{-5}$ error budget demanded by MUonE \cite{Banerjee:2020tdt}. Notice that there is no delicate cancellation in \cref{mu-pheno-est} in contrast to the accidental cancellation in \cref{nu-pheno-est}, and the binding correction is of natural size.

\section{Discussion and conclusions \label{sec:discussion}}
Using a simple perturbative analysis, the virial theorem, and the Koltun sum rule \cite{Koltun:1972kh}, we have identified a new model independent relationship between the expectation value of the single body potential operator, and binding corrections to high energy scattering. We have studied these corrections in detail for $\nu e\rightarrow \nu e$ and $\mu e \rightarrow \mu e$ scattering.  We find that due to rotational symmetry, corrections begin at $O(\epsilon_A/m_e)$ where $\epsilon_A$ is the binding energy of the target atom. Our main results are \cref{main-nue} for $\nu e \rightarrow \nu e$ scattering, and \cref{main-mue-a,main-mue-b} for $\mu e \rightarrow \mu e$ scattering. The results of this paper can also be applied to Moller scattering off atomic electrons orbiting hydrogen \cite{MOLLER:2014iki}. Since there are no $Z$-enhancements in hydrogen, we expect that the atomic binding corrections will be comparable to two-loop QED corrections and below the precision goals stated in Ref.~\cite{MOLLER:2014iki}.

In our analysis we have treated the final state electron as a free-particle solution (which amounts to using the impulse approximation). In general there will be perturbative corrections from Coulomb exchange with the final state system $\ket{B}$ that are {\it not included} in the radiative corrections to $\nu e \rightarrow \nu e $ and $\mu e \rightarrow \mu e$ scattering involving free electrons. These effects are well studied in e.g., non-relativistic contexts \cite{mott1965theory} but have not (to our knowledge) been computed for the ultra-relativistic kinematics we consider here. The  effect of final state interactions can be estimated in the case of hydrogen, but it is presently unclear if any universal form exists for many-body atoms. These corrections should be computed in the future and added to the binding corrections discussed in this paper (see Ref.~\cite{Plestid_Wise_2} for a discussion). 

The results presented here are important for ultra-precise measurements of muon electron scattering, as is planned for the MUonE experiment. Unlike in neutrino flux measurements, where the total cross section is most important, for MUonE the most relevant observable is the differential distribution with respect to the angles of outgoing muon and electron. The binding corrections discussed here will impact extractions of hadronic vacuum polarization from MUonE. Importantly, the error budget there is $\sim 10^{-5}$ and even for light nuclei (e.g., carbon with $Z^{4/3}\approx 11$) atomic binding corrections must be incorporated. 



\section*{Acknowledgements}
RP thanks Alexis Nikolakopoulos for helpful discussions. RP is supported by the Neutrino Theory Network under Award Number DEAC02-07CH11359. RP and MW are supported by the U.S. Department of Energy, Office of Science, Office of High Energy Physics under Award Number DE-SC0011632, and by the Walter Burke Institute for Theoretical Physics. 
\appendix

\section{Sum rules \label{app:sum-rules} }
Since the Koltun sum rule \cite{Koltun:1972kh} is very simple, but may be unfamiliar to some readers, we provide a self contained discussion (see also Ref.~\cite{Polls:1994zz}). Let us consider the hole spectral function for a state $\ket{A}$
\begin{align}
    S_A(\epsilon,\vb{p}) =& \int \frac{\dd^3 q}{(2\pi)^3} \sumint_B \mel{A}{\hat{a}^\dagger_{\vb{p}}}{B} \delta(\epsilon_A-\epsilon_B-\epsilon) \mel{B}{\hat{a}_{\vb{q}}}{A}~. 
\end{align}
The first (trivial) sum rule stems from the identity $ \int \dd \epsilon  \delta (x-\epsilon) f(x) = f(x)$ for any test function. Therefore, 
\begin{equation}
    \begin{split}
   \int \dd \epsilon S_A(\epsilon,\vb{p}) &= \int \frac{\dd^3 q}{(2\pi)^3} \bra{A}{\hat{a}^\dagger_{\vb{p}}} \sumint_B  \ketbra{B}{B}{\hat{a}_{\vb{q}} }\ket{A} ~\\
   &=  \int \frac{\dd^3 q}{(2\pi)^3} \mel{A}{   \hat{a}_{\vb{p}}^\dagger \hat{a}_{\vb{q}} }{A}~. 
   \end{split}
\end{equation}
When weighted against $\vb{p}^2/2m_e$ and integrated over $\dd^3p$ this gives $\langle \hat{T} \rangle$. 

Next, for the Koltun sum rule, we re-write $\epsilon$ in terms of the Hamiltonian via  (suppressing $\int \dd^3 q/(2\pi)^3$ for brevity's sake),
\begin{align}
    &~~~~ \epsilon \sumint_B \mel{A}{\hat{a}^\dagger_{\vb{q}}}{B} \delta(\epsilon_A-\epsilon_B-\epsilon) \mel{B}{\hat{a}_{\vb{p}} }{A} ~\\
    &= (\epsilon_A-\epsilon_B) \sumint_B \mel{A}{\hat{a}^\dagger_{\vb{q}}}{B} \delta(\epsilon_A-\epsilon_B-\epsilon) \mel{B}{\hat{a}_{\vb{q}} }{A} ~,\nonumber\\
    &= \sumint_B \mel{A}{\hat{a}^\dagger_{\vb{p}}}{B} \delta(\epsilon_A-\epsilon_B-\epsilon) \mel{B}{\hat{a}_{\vb{q}} \epsilon_A - \epsilon_B \hat{a}_{\vb{q}} }{A} ~,\nonumber \\
    &= \sumint_B \mel{A}{\hat{a}^\dagger_{\vb{p}}}{B} \delta(\epsilon_A-\epsilon_B-\epsilon) \mel{B}{\hat{a}^\dagger_{\vb{p}}[\hat{H},\hat{a}_{\vb{q}}] }{A}~.\nonumber
\end{align}
where we have used $\hat{H}\ket{A,B}= -\epsilon_{A,B} \ket{A,B}$. 
Then, using $(-\epsilon)$ instead of $\epsilon$ and integrating we obtain 
\begin{equation}
    \int \dd \epsilon~ (-\epsilon) S_A(\epsilon,\vb{p}) = \int \frac{\dd^3q}{(2\pi)^3} \mel{A}{\hat{a}^\dagger_{\vb{p}}[\hat{a}_{\vb{q}},\hat{H} ] }{A}~.
\end{equation}
Taking a Hamiltonian with one- and two-body operators, 
\begin{equation}
    \begin{split}
    \hat{H}= &\sum_{\vb{p}} \frac{\vb{p}^2}{2m_e}\hat{a}_{\vb{p}}^\dagger\hat{a}_{\vb{p}}  + \sum_{\vb{p},\vb{p}'} V_1(\vb{p},\vb{p}') \hat{a}_{\vb{p}'}^\dagger\hat{a}_{\vb{p}}\\
    &~~~~+  \sum_{\vb{p},\vb{p}',\vb{q},\vb{q}'} V_2(\vb{p},\vb{p}',\vb{q},\vb{q}') \hat{a}_{\vb{p}'}^\dagger\hat{a}_{\vb{q}'}^\dagger\hat{a}_{\vb{p}} \hat{a}_{\vb{q}}~,
    \end{split}
\end{equation}
evaluating the commutator, and integrating over $\dd \epsilon$ to remove the delta function, we arrive at
\begin{equation}
    \hspace{-0.01\linewidth} \int \frac{\dd^3p}{(2\pi)^3} \int\!\dd \epsilon(-\epsilon )S_A(\epsilon,\vb{p}) = \frac{\mel{A}{\hat{T} + \hat{V}_1 + 2 \hat{V}_2}{A}}{\braket{A}{A}}.
\end{equation}
For a more general Hamiltonian written as an expansion $n$-body operators as $\hat{H}= \sum_n \hat{\mathcal{O}}_n$ one finds \cite{Koltun:1972kh}
\begin{equation}
     \int \frac{\dd^3p}{(2\pi)^3} \int \dd \epsilon (-\epsilon)
     S_A(\epsilon,\vb{p}) = \sum_n  n \frac{ \mel{A}{ \hat{\cal O}_n}{A}}{\braket{A}{A}}~.
\end{equation}
\section{Thomas-Fermi estimate \label{TF-model}}
The Thomas-Fermi model offers a crude description of bulk atomic properties \cite{March2016self}. Since we have reduced the problem of binding corrections to evaluating $\langle \hat{V}_1 \rangle_A$ the quality of the estimate depends only on the accuracy with which we can model $n_A(\vb{r})$. 

The Thomas-Fermi model can be summarized as $ r^2 n(\vb{r}) \propto x^{1/2} [f(x)]^{3/2}$ where $x=r/b_0$ with $b_0 = 0.88 Z^{-1/3} a_0$. The function $f(x)$ is determined by solving the differential equation, 
\begin{equation}
    x f''(x) = x^{1/2} [f(x)]^{3/2}~,
\end{equation}
with $f(0)= 1$ and $f'(0)\simeq 1.588$.  Using this solution, and integrating numerically, one finds 
\begin{equation}
    \frac{\int \dd x ~ \qty(\frac{1}{x}) x^{1/2}f(x)}{\int \dd x ~ x^{1/2}f(x)} \approx 1.79~.  
\end{equation}
Converting to physical units gives 
\cref{thomas-fermi-est}. 

\section{Relativistic corrections \label{app:rel-corr} }
In \cref{hyd-exp} we have written a hydrogen (or hydrogen-like) atom in terms of an entangled proton (nucleus) and a free-electron state. In the limit of $m_p/m_e\rightarrow \infty$ the proton (nucleus) acts as a static Coulomb field and the exact solution (to all orders in $Z\alpha$) is given by the solution of the Dirac equation in a central potential. These wavefunctions generically {\it cannot} be written in the form of \cref{hyd-exp}. Nevertheless, as we dicuss below, these effects begin first at $O(p^3/m_e^3)$ and are therefore sub-dominant in our expansion. 

Consider the Dirac equation in a central potential. Using the ``Bjorken and Drell'' basis this can be written as 
\begin{equation}
    \begin{pmatrix}
        \phantom{2m}-\epsilon_H + V(\vb{x}) & -\iu \vb*{\sigma} \cdot \vb*{\nabla} \\
        \iu \vb*{\sigma} \cdot \vb*{\nabla} & 2m -\epsilon_H + V(\vb{x}) 
    \end{pmatrix}
    \begin{pmatrix}
        U(\vb{x}) \\
        L(\vb{x})
    \end{pmatrix}
    =0 ~.
\end{equation}
Non-relativistic bound states satisfy the counting $\vb*{\nabla} \sim O(\lambda)$, $V(\vb{x})\sim O(\lambda^2)$, and $\epsilon_H \sim O(\lambda^2)$, as required by the virial theorem. We may then expand order-by-order in $\lambda$,
\begin{align}
    U(\vb{x}) &= U^{(0)}(\vb{x}) + U^{(2)}(\vb{x}) + U^{(4)}(\vb{x}) + \ldots~,\\
    L(\vb{x}) &= L^{(1)}(\vb{x}) + L^{(3)}(\vb{x}) + L^{(5)}(\vb{x})  + \ldots ~,\\
    \epsilon_H&= \epsilon_H^{(2)} + \epsilon_H^{(4)} + \epsilon_H^{(6)} + \ldots ~,
\end{align}
where we have anticipated the vanishing of alternating orders. The fact that $\vb*{\nabla}$ is the dominant term in the expansion implies the constraint at leading order, 
\begin{equation}
    \label{constraint}
    L^{(1)}(\vb{x}) = \frac{-\iu \vb*{\sigma}\cdot \vb*{\nabla}}{2m} U^{(0)}(\vb{x}) ~. 
\end{equation}
The zeroth order upper component satisfies the Schr\"odinger equation, 
\begin{equation}
   \qty[ -\frac{1}{2m}\nabla^2 + V(\vb{x}) ] U^{(0)}(\vb{x})= -\epsilon_H^{(0)} U^{(0)}(\vb{x})~.
\end{equation}
The second order correction $U^{(2)}(\vb{x})$ incorporates fine-structure effects including, for example, spin-orbit coupling.
We may therefore, through $O(\lambda^2)=O(\epsilon_H/m_e)$, write the hydrogen-like Dirac wavefunction as\footnote{This result  generalizes Problem \S 39 and Eq.\ (57.3) of \cite{Berestetskii:1982qgu} to $O(v^2/c^2)$ in the non-relativistic expansion.},
\begin{equation}
    \label{dirac-spinor-x}
    \Psi(\vb{x}) =  \begin{pmatrix}
                            \psi_H(\vb{x}) \\
                            \frac{-\iu \vb*{\sigma}\cdot \vb*{\nabla}}{2m_e} \psi_H(\vb{x}) 
                     \end{pmatrix}  + O(\lambda^3)~,
\end{equation}
where $\psi_H(\vb{x)} \equiv U^{(0)}(\vb{x}) + U^{(2)}(\vb{x})$ is a two-component spinor. Note that in the spirit of degenerate perturbation theory, there always exists a basis such that the upper component spinor in \cref{dirac-spinor-x} is an eigenstate of the spin-orbit coupling $\vb{L}\cdot \vb{S}$. 

Equivalently in momentum space we find 
\begin{equation}
    \label{dirac-spinor-p}
    \tilde{\Psi}(\vb{p}) =  
    \begin{pmatrix}
            \tilde{\psi}_H(\vb{p})  \\
                \frac{\vb*{\sigma}\cdot \vb{p}}{2m_e} \tilde{\psi}_H(\vb{p})  
    \end{pmatrix} + O(\lambda^3)~.
\end{equation}
Note in \cref{dirac-spinor-x}  we are not assuming that ${\tilde \psi}({\bf p})$ is a rotationally invariant function of the three momentum.  Hence this equation holds not just for the ground state, but orbitally excited states as well. 
We see that up to a normalization of states (i.e., relativistic vs. non-relativistic conventions) this is equivalent to using \cref{hyd-exp} up to $O(\lambda^3)$ corrections. 

When computing corrections to the matrix element, the contributions from $U^{(2)}$ enter only via $\int \dd^3p/(2\pi)^3  |\tilde{\psi}_H(\vb{p})|^2 |{\sf M}_0|^2$. Since ${\sf M}_0$ is independent of $\vb{p}$ (by definition) there is no correction, because the norm of the wavefunction is fixed to all orders $\int \dd^3p/(2\pi)^3 |\tilde{\psi}(\vb{p})|^2=1$. When computing the terms proportional to $\vb{p}^2/m_e^2$ and $\epsilon/m_e$ in e.g., \cref{Hyd-sigma} it is sufficient to use the zeroth-order approximation to the wavefunction.  

The above analysis makes it clear that this approximations (likely) holds to the same accuracy in many-body atoms. One can write the full QED Hamiltonian, and similarly expand the result order-by-order in $\lambda$. The constraint \cref{constraint} will appear independent of the details of the interactions between electrons, all of which are counted as $O(\lambda^2)$. 

The fully relativistic spectral function  satisfies the exact sum rule $\int \dd^3 p/(2\pi)^3 \int \dd \epsilon ~S(\epsilon, \vb{p}) = Z$. When computing corrections due to bound states, i.e., for terms proportional to $\vb{p}^2/m_e^2$ or $\epsilon/m_e$, it suffices to use the non-relativistic approximation of the spectral function.

\bibliographystyle{apsrev4-1}
\bibliography{biblio}

\end{document}